\documentclass[runningheads]{llncs}
\usepackage[T1]{fontenc}
\usepackage{graphicx}
\usepackage{hyperref}
\usepackage{amsmath}
\usepackage{amssymb}
\usepackage{mathtools}
\usepackage{booktabs}
\usepackage{graphicx}

\usepackage{color}

\usepackage{makecell}

\begin{document}

\title{SimpliPy: A Source-Tracking Notional Machine for Simplified Python}
%
%
\author{\makebox[.5\textwidth][c]{Moida Praneeth Jain\orcidID{0009-0006-8192-6041} \and Venkatesh Choppella\orcidID{0000-0003-1085-3464}}}

\authorrunning{M. P. Jain and V. Choppella}

\institute{International Institute of Information Technology Hyderabad, Telangana, India\\
	\email{moida.praneeth@students.iiit.ac.in}, \email{venkatesh.choppella@iiit.ac.in}}

\maketitle

\begin{abstract}
	Misconceptions about program execution hinder many novice programmers. We
	introduce SimpliPy, a notional machine designed around a carefully chosen Python
	subset to clarify core control flow and scoping concepts. Its foundation is a
	precise operational semantics that explicitly tracks source code line numbers
	for each execution step, making the link between code and behavior unambiguous.
	Complementing the dynamic semantics, SimpliPy uses static analysis to generate
	Control Flow Graphs (CFGs) and identify lexical scopes, helping students build a
	structural understanding before tracing. We also present an interactive
	web-based debugger built on these principles. This tool embodies the formal
	techniques, visualizing the operational state (environments, stack) and using
	the static CFG to animate control flow directly on the graph during step-by-step
	execution. SimpliPy thus integrates formal semantics, program analysis, and
	visualization to offer both a pedagogical approach and a practical demonstration
	of applying formal methods to program understanding.

\end{abstract}

\section{Motivation and Introduction}

\subsection{Misconceptions and Difficulties in Learning
	Programming}

Learning programming encompasses the acquisition of a
multitude of related skills: reading programs, tracing their
execution, debugging, writing test cases, documenting and
writing code. Acquiring these skills by students is
hampered by the many misconceptions they harbour about how
programs are structured and how they run
\cite{2020-johnson-cep-py-misconceptions,2010-kaczmarczyck-sigcsets-misconceptions,2024-lu-acm-proceedings-pl-misconceptions,2007-vainio-iticse-tracing-skills}.

In this paper, we are motivated by the larger problem of developing a pedagogy
of teaching how programs run.  Students construct a variety of mental models and
often represent them as
`sketches'~\cite{2017-cunningham-icer-tracing-replication-study} when tasked with
the problem of tracing code. These sketches reflect the thought processes and
mental models that students construct to express their understanding of program
execution.

\subsection{Syntactic vs. Semantic Approach}
In the teaching of programming, we are seeing a shift in pedagogy that
increasingly emphasizes a semantic, rather than a syntactic approach to
understanding
programs~\cite{2019-guzdial-dagstuhl-notional-machine-semantics,2013-sorva-toce-notional-machines-survey}.
The semantic view emphasizes program
comprehension~\cite{2017-nelson-icer-comprehension-first}, encourages the
construction of mental models that reflect an accurate understanding of how a
program runs on a \emph{notional
	machine}~\cite{1981-boulay-ijmms-notional-machine}. Understanding the notional
machine and its dynamics then becomes central to understanding how any program
in a programming language runs. Indeed, it has been argued that teaching the
notional machine ought to be an explicit objective in introductory programming
education~\cite{2013-sorva-toce-notional-machines-survey}.  An accurate
definition of a notional machine necessarily borrows from the theory of
programming language semantics.  The effective teaching and learning of
programming is therefore intimately intertwined with the knowledge of the
semantics of the programming language of the program
\cite{2008-reynolds-sigplan-notices-teaching-programming}.

\subsection{Purpose and specificity of a notional machine}

Knowledge of a programming language spans multiple
dimensions, each of which needs to be understood to gain
mastery of the language. The \textit{control} dimension specifies
how the program's flow moves across different program
locations. The \textit{scope} dimension specifies how variables are
looked up. The \textit{datatype} dimension specifies how data is
defined. The \textit{storage} dimension specifies how data is laid
out in memory and shared. When designing notional machines,
one needs to consider which knowledge dimensions of the
language are being
modeled~\cite{2019-guzdial-dagstuhl-notional-machine-semantics}.
A notional machine that tries to address all knowledge
dimensions might end up being too complicated for the
student to follow. It therefore typically focuses on one or
more, but not necessarily all knowledge
dimensions~\cite{2020-iticse-working-group-notional-machines,2013-sorva-toce-notional-machines-survey}.
Its purpose then becomes the need to clearly elucidate the
program's behaviour along the collection of dimensions.

A pedagogy based on semantics should encourage constructing
well-defined artefacts around the program's structure that
also approximate its putative runtime behaviour. These
artefacts should be unambiguously interpretable to relate to
the program's actual behaviour along a given set of
knowledge dimensions. The artefacts could be diagrams,
formula, or even plain text. Examples of such artefacts are
stack diagrams for procedure calls, box and pointer diagrams
for structure sharing, labelled directed graphs for control
flow~\cite{2020-iticse-working-group-notional-machines}.

\subsection{SimpliPy}

Towards this end, we introduce \textbf{SimpliPy}, a new notional machine
designed specially to study control flow and scoping in Python. The SimpliPy
notional machine operates on a subset of Python, suitably chosen to make tracing
easier, while retaining the essential control features of the language.

Python has wide usage as the programming language of choice for introductory
programming in many parts of the
world~\cite{2024-mason-sigcsets-prog-lang-survey}. Therefore, our focus on
Python seems well justified.

The SimpliPy semantics is precise, yet accessible after some careful practice.
SimpliPy exploits the line-oriented structure of Python, enabling precise source
tracking, meaning it can trace the origin of every transition back to a specific
program location in the source code. This level of detail allows students to
connect each operational step directly to the Python code they see, reinforcing
their understanding of how specific lines influence program behavior.

SimpliPy assumes students of programming are familiar with the elementary
concepts of sets, relations, (partial) functions, and sequences, usually taught
in high-school.

\section{Related Work}

We present a brief tour of related work and connect it with the SimpliPy
pedagogy. The concept of a notional machine goes back to over forty years when
Boulay~et~al.~\cite{1981-boulay-ijmms-notional-machine} introduced the idea of a
notional machine as ``an idealised, conceptual computer whose properties are
implied by the constructs in the programming language employed''.
Sorva~\cite{2013-sorva-toce-notional-machines-survey} emphasizes that a notional
machine should be explicitly studied as part of programming education. The
SimpliPy approach embraces this pedagogical principle.

Various notional machines, their evolution and uses have been explored by the
ITiCSE 2020 Working Group on Notional
Machines~\cite{2020-iticse-working-group-notional-machines}. However, the
emphasis of these notional machines discussed is visual and informal. The
SimpliPy notional machine, on the other hand, is based on a precise mathematical
semantics of a well-defined subset of Python. Understanding the SimpliPy
semantics, however, only demands a facility with elementary set theory. SimpliPy
semantics can be stated (a bit more verbosely) in natural language. This makes
it accessible to beginning students of programming.

Dickson et al.~\cite{2020-dickson-iticse-rise-of-notional-machines} argue that
the complexity of notional machines should increase as the teaching goes from
the early to advanced stages. The various types of control flow in Python:
sequential, conditional, iterative and procedural are dealt with in an
incremental way in SimpliPy. Nelson and
co-workers~\cite{2017-nelson-icer-comprehension-first,2018-xie-sigcsets-scaffolding}
advocate a ``comprehension first'' approach to programming. They emphasise that
understanding the structure and dynamics of a program should be preceded by
first introducing an appropriate ontology of concepts. In their work, the
semantics of a program is traced to the control flow path in the code of the
underlying interpreter.

SimpliPy introduces its notional machine using the terminology of systems
thinking (observables, states, actions, transitions) and it introduces the key
notion of closures, crucial to understanding functions.

The SimpliPy approach shares the spirit of
Nelson~et~al.~\cite{2017-nelson-icer-comprehension-first} in using interpreter
rules to define program flow, but differs in three key aspects. First,
SimpliPy's notional machine instructions directly correspond to the program's
source code instructions, allowing students to apply the machine's rules by
interpreting the code they see. Second, the SimpliPy pedagogy encourages initial
understanding through paper-and-pencil exercises – constructing execution traces
and state diagrams by hand leverages the cognitive benefits of slower,
deliberate engagement~\cite{2014-mueller-psy-science-pen-vs-keyboard}. This
manual process is then actively supported and reinforced by an interactive
visualization tool, which dynamically displays the same execution states
(environments, stack) and control flow transitions, allowing students to verify
their understanding and explore execution step-by-step. Third, SimpliPy promotes
a comprehension-first strategy by requiring students to construct intermediate
artifacts derived from static analysis before tracing a program. These artifacts
include scope-related information (lexical blocks, variable declarations) and
control information (control transfer functions, the program's control flow
graph), which are also visualized by our tool to solidify the link between
static program structure and dynamic behavior.

Notional machines developed for Python in the past have been either informal and
incomplete~\cite{2022-dickson-sigcsets-notional-machine-classroom} or have
addressed only very specific issues like mutation and
sharing~\cite{2020-johnson-cep-py-misconceptions} or have focused on identifying
a subset for better error messages~\cite{2023-aycock-wscce-spy3}. Lu and
Krishnamurthi~\cite{2024-lu-acm-proceedings-pl-misconceptions} present an online
platform called SMoL Tutor. It comes with a web application called Stacker that
steps through a program's execution. SMoL handles multiple languages, including
Python. However, it requires that code be edited in a Lisp-style syntax and
precludes experimenting with scoping mechanisms unique to Python, like
\texttt{nonlocal}, \texttt{global} and function-only scope. On the other hand,
in SimpliPy, all functions need to be named (the \texttt{lambda} keyword is not
considered, but this is not a serious limitation because a fresh name can always
be introduced).

Existing formal semantics for Python lack source tracking, limiting their
utility in educational contexts. Smeding~\cite{2009-smeding-thesis-minpy} models
Python 2.5 as a transition system and does track instructions, but the approach
omits some scoping details (like the global keyword) and is outdated for modern
Python versions. Politz et al.~\cite{2013-politz-oopsla-lambdapy} rewrite Python
to a scheme-like core language, which introduces scoping inaccuracies with
classes, while Guth~\cite{2013-guth-thesis-kpython} and
Köhl~\cite{2021-kohl-thesis-term-unification} employ rewrite-based models that do
not directly follow Python source lines. These limitations contrast with
SimpliPy, which preserves source-level tracking, making it a more effective tool
for teaching Python’s execution flow.

\section{SimpliPy Overview}

In this section, we provide a comprehensive overview of the SimpliPy notional
machine, detailing its language subset, the structure of the programs, the
static analysis artefacts it utilizes, the notional machine's state, and the
dynamics governing its execution.

\subsection{Language Subset and Program Structure}
SimpliPy operates on a carefully curated subset of Python designed for teaching control flow and scoping. It follows Python's line-oriented structure but introduces specific constraints for pedagogical purposes.

\subsubsection{Main Syntactic Entities}
\begin{itemize}
	\item \textbf{Expressions (\texttt{Exp}):} Simple computations including
	      constants, variables and operations using unary or binary operators
	      (\texttt{Uop}, \texttt{Bop}). Crucially, expressions in SimpliPy are designed to
	      be simple and \textit{do not} contain function calls.
	\item \textbf{Instructions (\texttt{Instr}):} The basic syntactic unit
	      corresponding to a single line of Python code within the SimpliPy subset.
	\item \textbf{Statements (\texttt{Stmt}):} Represent complete logical steps
	      or control structures. A simple statement might correspond directly to a single
	      instruction (like \texttt{pass} or \texttt{ExpAssign}). Compound statements
	      (\texttt{If}, \texttt{While}, \texttt{Def}) encompass multiple instructions,
	      including nested blocks.
	\item \textbf{Blocks (\texttt{Blk}):} A sequential composition of one or
	      more statements. Blocks may define scopes and group related statements, such as
	      the body of an \texttt{if}, \texttt{while}, or \texttt{def}.
	\item \textbf{Program (\texttt{Pgm}):} The entire SimpliPy program, defined
	      as a single top-level block.
\end{itemize}

A SimpliPy program $P$ of length $N$ (containing $N$ instructions across all
lines) can be viewed as a mapping from locations (line numbers) to instructions:
\[ P: [1..N] \rightarrow \texttt{Instr} \] We define the set of program
locations as $L = [1..N+1]$, where $N+1$ represents a conceptual end-of-program
location.

The precise syntactic structure allowed in SimpliPy is formally defined by a BNF
grammar, which is available in our supplementary materials repository:
\url{https://github.com/PraneethJain/simplipy}.

\subsection{Static Analysis}
An important aspect of the pedagogy of SimpliPy is the analysis of a program
before it is run. This involves two types of analyses: scope and control, each
producing multiple artefacts. Scope analysis defines the program’s lexical
blocks, identifies \emph{locals} (the set of variables declared within each
lexical block, excluding nonlocal and global variables), \emph{nonlocals} and
\emph{globals} (variables designated as nonlocal or global within each block).
Control flow analysis generates three primary artifacts: a structural
abstraction of the program where each statement is represented by its syntactic
category, a set of control transfer functions utilized by the notional machine’s
transition function, and a control flow graph that statically approximates the
trace of program locations during execution.  Students are expected to construct
these artefacts as part of exercises in program comprehension before they build
the trace of the program.

The control transfer functions are partial functions that map a location to
another location. SimpliPy consists of four control transfer functions:
\textit{next}, \textit{true}, \textit{false}, and \textit{err}. The functions
\textit{true} and \textit{false} are specifically associated with if and while
statements, determining the control flow based on the evaluated condition. The
\textit{err} function is defined universally for all locations. \textit{next}
facilitates sequential execution, transitioning to the location corresponding
to the subsequent statement in the program, while \textit{err} enables handling
of erroneous cases, mapping a location to itself, to indicate a fixed point.

Additionally, \textit{next} is configured to manage control flow for break and
continue statements appropriately. For instance, break within a while loop
redirects control to the location mapped by the \textit{false} function of the
while statement, thereby exiting the loop. Conversely, continue within a while
loop directs control back to the location of the while statement itself,
enabling a re-evaluation of the loop condition.

\subsection{Notional Machine Structure}

\begin{displaymath}
	\begin{aligned}
		                  & (e, h, k) \in State                         \\
		State= LexicalMap & \times LexicalHierarchy \times Continuation
	\end{aligned}
\end{displaymath}

The notional machine is a labelled transition system whose state space is
defined by the triple consisting of the following: the lexical map $e$, the
lexical hierarchy $h$, and the continuation $k$.

\begin{itemize}
	\item \textbf{Lexical Map ($e$):} This component holds all the environments
	      created during execution. It functions as a map from unique environment
	      identifiers (integers, with 0 representing the global environment) to
	      environments. An environment itself is a dictionary mapping variable names
	      (identifiers) within that scope to their currently assigned values. Values can
	      include primitive types (numbers, booleans, strings), special markers (like an
	      uninitialized `bottom' value, denoted $\bot$), or closures representing function
	      definitions. The lexical map is essentially an abstraction of the program's
	      memory for variable bindings across all scopes.

	\item \textbf{Lexical Hierarchy ($h$):} This structure captures the well
	      founded relation on the environments. It maps each environment's unique
	      identifier (except the root) to the identifier of its immediate parent
	      environment. The global environment (Id 0) serves as the root of this tree and
	      has no parent. This hierarchy is fundamental to understanding Python's lexical
	      scoping: when resolving a variable name, a typical search begins in the current
	      environment and proceeds upwards towards the root by following the parent links
	      defined in $h$ until the variable is found or the global scope is checked. New
	      branches in the tree are created during function calls, linking the newly
	      created function's environment to the environment where the function was
	      lexically defined.

	\item \textbf{Continuation ($k$):} This component is an abstraction of the
	      control state, acting as the machine's execution stack. It comprises a list of
	      \textit{contexts}. A context is defined as a tuple (location, environment\_id),
	      indicating the instruction to be executed (location) and the environment
	      (environment\_id) required for its execution. The context at the head of the list
	      (top of the stack) determines the immediate step. Function calls introduce a new
	      context onto the list, preserving the caller's return point (location of the
	      call, caller's environment Id). Function returns remove the current context,
	      effectively transferring control back according to the previously preserved
	      context. In this way, the continuation directs the control flow through
	      procedure calls and returns.

\end{itemize}

Initially, the notional machine is configured as follows: The lexical map $e$
contains a single entry for the global environment, identified by Id 0. The
lexical hierarchy $h$ contains only the global environment Id 0, which has no
parent. The continuation $k$ begins with one context on its stack, pointing to
the location of the program's first instruction and the global environment Id 0.

\subsection{Notional Machine Dynamics}

The notional machine is parametrized by the program $P$, which maps locations to
instructions. The actions on the notional machine are program instructions. The dynamics of the notional machine is specified by a transition relation

\begin{displaymath}
	\begin{aligned}
		(e, h, k) \xrightarrow{c}{} (e', h', k')
	\end{aligned}
\end{displaymath}

where $c=P_{i}$ is the instruction at the current location $i$ from the top context.
\subsubsection{Variable Lookup Mechanism}
Before detailing the state transitions, it is essential to understand how SimpliPy resolves variable names, adhering to Python's lexical scoping rules. When the value of a variable \texttt{var} is needed within the current execution context (defined by environment \texttt{env\_id}), the lookup proceeds as follows:

\begin{enumerate}
	\item \textbf{Current Environment:} The environment identified by
	      \texttt{env\_id} is checked first. If \texttt{var} is found here, its value is
	      returned.

	\item \textbf{Lexical Ancestors:} If \texttt{var} is not in the current
	      environment, the search continues recursively in the parent environment,
	      determined by consulting the lexical hierarchy ($h$). This process follows the
	      chain of parent links upwards from the current environment.

	\item \textbf{Global Environment:} If the search reaches the global
	      environment (Id 0) without finding \texttt{var} locally or in any ancestor
	      scope, the global environment is checked.

	\item \textbf{Nonlocal/Global Directives:} The static analysis identifying
	      \texttt{nonlocal} and \texttt{global} variables within lexical blocks influences
	      this search. If a variable is marked \texttt{nonlocal}, the search explicitly
	      skips the immediate local environment and the global environment, and starts in
	      the parent environment. If marked \texttt{global} (or if it's accessed in the
	      top-level scope), the search targets the global environment directly or proceeds
	      there if not found locally.

	\item \textbf{Lookup Failure:} If the variable is not found after checking all
	      relevant lexical ancestors, a lookup error occurs.
\end{enumerate}

The update mechanism for assignments similarly uses this lookup process to find
the correct environment where an existing variable binding should be updated, or
determines that a new binding needs to be created (the global scope for
top-level assignments or variables declared \texttt{global}).

The precise mathematical rules defining the transition relations for each
instruction type are detailed in the formal operational semantics, available in
our supplementary materials repository:
\url{https://github.com/PraneethJain/simplipy}. We briefly describe the
conceptual behavior for each instruction type below.

\begin{description}

	\item[Pass, Global, Nonlocal:] These instructions primarily serve structural
		or declarative purposes. \texttt{Pass} performs no operation. \texttt{Global}
		and \texttt{Nonlocal} affect the static analysis and subsequent variable lookups
		but do not modify the machine state during their own execution step. Control
		simply proceeds to the next instruction determined by the \texttt{next} control
		transfer function.

	\item[Break, Continue:] These alter control flow within loops.  \texttt{Break}
		transfers control to the instruction immediately following the innermost
		enclosing \texttt{while} loop.  \texttt{Continue} transfers control back to the
		condition check (the \texttt{while} instruction itself) of the innermost
		enclosing loop. The \texttt{next} control transfer function determines this
		behaviour.

	\item[Expression Assignment:] The expression on the right-hand side is
		evaluated using the current environment context. If successful, the variable on
		the left-hand side is updated with the resulting value in the appropriate
		environment (determined by the lookup mechanism). Control then moves to the next
		instruction. Evaluation errors lead to an error transition.

	\item[If / While:] The conditional expression is evaluated. Based on whether
		the result is true or false, control transfers to the location specified by the
		corresponding \texttt{true} or \texttt{false} control transfer function
		(entering the appropriate block for \texttt{if}, entering or exiting the loop
		for \texttt{while}).  Non-boolean results or evaluation errors lead to an error
		transition.

	\item[Function Definition:] A closure value is created, capturing the starting
		location of the function's code block, the list of formal parameter names, and a
		reference to the current lexical environment (its Id). This closure is then
		bound to the function's name in the current environment. Control proceeds to the
		next instruction.

	\item[Call Assignment:] The expression identifying the function is evaluated
		to retrieve a closure. The argument expressions are evaluated. If successful and
		type/arity checks pass, a new environment is created for the function call.
		Formal parameters are bound to the evaluated argument values in this new
		environment, and other local variables declared within the function are
		initialized to $\bot$. The new environment is linked to its parent (based on the
		closure's captured environment Id) in the lexical hierarchy ($h$).  The context
		for the function's entry point and the new environment Id is pushed onto the
		continuation $k$. Control transfers to the function's entry point. Errors during
		lookup or evaluation lead to an error transition.

	\item[Return:] The return expression is evaluated in the function's current
		environment. The top context is popped from the continuation stack (the
		function's current context). The variable targeted by the original call
		assignment instruction (found at the top of the stack) is updated in the
		caller's environment with the evaluated return value. Execution resumes at the
		instruction following the call site in the caller's context. Evaluation errors
		lead to an error transition.

\end{description}

An execution trace consists of the sequence of states generated by repeatedly
applying these transition rules, starting from the initial state, until a final
state (fixed point) is achieved.

\section{SimpliPy Visualization Tool}

\begin{figure}[htbp]
	\centering
	\includegraphics[width=\textwidth]{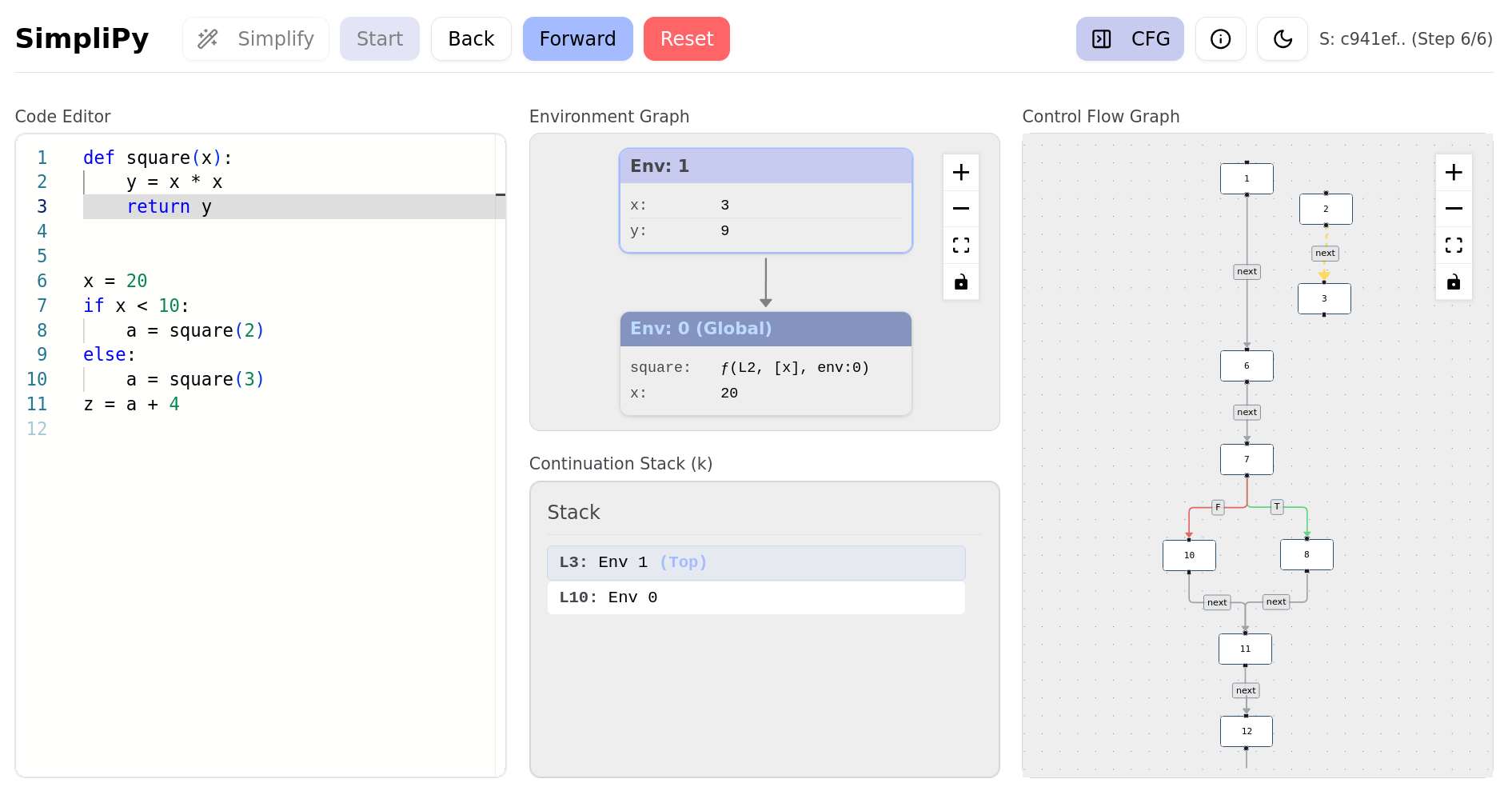}
	\caption{The SimpliPy interactive visualization tool interface. Coordinated
		views display the program state according to the SimpliPy notional machine:
		(Left) Source code with the current instruction highlighted. (Top
		Middle) The Environment Scope View visualizes the lexical map ($e$) and
		hierarchy ($h$) as a tree, showing active environments  and their variable
		bindings. (Bottom Middle) The Continuation Stack ($k$) displays active execution
		contexts $(location, env\_id)$. (Right) The Control Flow Graph (CFG) highlights
		the edge representing the most recent control transfer. Debugging controls at
		the top allow simplifying code, starting/resetting execution, stepping
		forward/backward, and toggling the CFG view.}
	\label{fig:simplipy_tool_interface}
\end{figure}

To operationalize the SimpliPy notional machine and support student learning,
we developed an interactive web-based visualization tool, publicly accessible at \url{https://simplipy.web.app/}.
This tool serves as a dynamic complement to manual program tracing, allowing students to observe the
step-by-step execution according to the defined semantics, verify their
understanding, and explore the relationship between code, state, and control
flow. The design prioritizes pedagogical clarity and faithful semantic
representation over the feature breadth of production development tools.

\subsection{Pedagogical Features}

The SimpliPy tool presents coordinated views designed to showcase concepts
central to the notional machine:

\begin{itemize}

	\item \textbf{Source Code View:} Displays the Python code, persistently
	      highlighting the current line number indicated by the top of the
	      continuation stack. This directly links the machine's current focus to the
	      corresponding source instruction.

	\item \textbf{Environment Scope View:} A key feature that explicitly
	      visualizes the relationship between different scopes ($e$ and $h$) active during
	      execution, rendered as a tree. Each node represents an environment and displays
	      the variables currently bound within that specific scope along with their values
	      (including closures or uninitialized markers $\bot$). This makes Python's
	      scoping rules tangible, allowing students to see exactly which environment holds
	      a given variable and understand the lookup process.

	\item \textbf{Continuation Stack View:} Displays the current continuation
	      ($k$), showing the sequence of active execution contexts \texttt{(location,
		      env\_id)}. This clarifies procedural control flow, showing pending return points
	      and their associated environments.

	\item \textbf{Control Flow Graph View:} Integrates the static control flow
	      graph.  During execution stepping, the tool highlights the edge representing the
	      control transfer just taken (e.g., `next', `true', `false'), visually connecting
	      the dynamic execution path to the program's static structure. The `err' control
	      transfer function is omitted from this view to reduce clutter.

	\item \textbf{Step-by-Step Execution Control:} Enables forward execution one
	      instruction at a time, following the SimpliPy transition rules precisely. A
	      history mechanism allows stepping backward to review previous states.

	\item \textbf{Simplification Support:} An integrated feature attempts to
	      automatically convert standard Python code into the SimpliPy subset, lowering
	      the barrier for using the tool with existing examples.

\end{itemize}

\subsection{Comparison with Other Tools}

The SimpliPy tool occupies a specific pedagogical niche, differing significantly
from both production debuggers and other educational or analysis tools,
primarily through its chosen level of abstraction and its direct embodiment of a
formal notional machine.

Compared to production debuggers found in environments like VS Code or PyCharm,
SimpliPy offers distinct pedagogical advantages via deliberate simplification.
Production tools, designed for development efficiency, often present complex
interfaces and expose low-level implementation details (e.g., specific stack
frame layouts, memory addresses) that can distract learners from core language
semantics. Their typical "Locals" view often flattens the representation of
nested scopes, obscuring the hierarchical nature of lexical environments crucial
for understanding variable lookup rules and lifetimes. While powerful for
debugging with features like breakpoints and live value modification, their
focus is not on illustrating the fundamental semantic steps in a clear,
abstracted manner. SimpliPy also allows stepping backward through the execution
history to review previous states and transitions, a feature generally absent in
the standard execution modes of production debuggers, which facilitates learning
and exploration of cause-and-effect in program execution. SimpliPy intentionally
abstracts away memory implementation specifics, focusing instead on a faithful,
observable step-by-step execution of its formal semantics, prioritizing
conceptual clarity.

SimpliPy also distinguishes itself from other educational and analysis tools.
While Python Tutor \cite{2013-guo-sigcse-python-tutor} provides excellent
visualizations of program state including memory layout (stack/heap), SimpliPy
deliberately abstracts these aspects to emphasize the lexical environment
hierarchy. Beginner-focused IDEs like Thonny \cite{2015-annamaa-koli-thonny}
offer a simplified debugging experience but operate closer to standard Python
execution. Tools like Localizer \cite{2024-khan-debt-localizer}, on the other
hand, focus on fault localization in full Python code using dynamic analysis of
test results and visualizing Control Flow Graphs (CFGs) annotated with
suspiciousness scores. In contrast to these tools, SimpliPy's unique
contribution lies in its specific focus on visualizing its formal operational
semantics for a simplified language subset—particularly scope structure (via the
environment tree) and control flow (via the continuation stack and CFG path
highlighting)—at a high level of abstraction, rigorously embodying its notional
machine and supporting both forward and backward stepping to maximize learning
impact. Table~\ref{tab:feature-comparison} provides a comparative summary of
these features.

\renewcommand{\arraystretch}{2.0}
\begin{table}[ht]
	\centering
	\caption{Pedagogical Feature Comparison: SimpliPy vs. Other Tools}
	\label{tab:feature-comparison}
	\setlength{\tabcolsep}{0pt}
	\begin{tabular}{l p{2cm} p{2cm} p{2cm} p{2cm} p{2cm}}
		\toprule
		\makecell{Feature}     &
		\makecell{SimpliPy       \\Tool} &
		\makecell{Localizer      \\\cite{2024-khan-debt-localizer}} &
		\makecell{Python Tutor   \\\cite{2013-guo-sigcse-python-tutor}} &
		\makecell{Thonny         \\\cite{2015-annamaa-koli-thonny}} &
		\makecell{VS Code        \\/ PyCharm} \\
		\midrule
		\makecell{Scope          \\Structure} &
		\makecell{Yes            \\(Tree)} &
		\makecell{No}          &
		\makecell{Limited        \\(Stack/Heap)} &
		\makecell{Simplified     \\Locals} &
		\makecell{No             \\(Flattened)} \\

		\makecell{Language       \\Subset} &
		\makecell{Restricted}  &
		\makecell{Full           \\Language} &
		\makecell{Broad          \\Subset} &
		\makecell{Full           \\Language} &
		\makecell{Full           \\Language} \\

		\makecell{CFG}         &
		\makecell{Yes}         &
		\makecell{Yes}         &
		\makecell{No}          &
		\makecell{No}          &
		\makecell{No}            \\

		\makecell{Variable       \\Modification} &
		\makecell{No}          &
		\makecell{No}          &
		\makecell{No}          &
		\makecell{Yes}         &
		\makecell{Yes}           \\

		\makecell{Breakpoints} &
		\makecell{No}          &
		\makecell{No}          &
		\makecell{Limited}     &
		\makecell{Yes}         &
		\makecell{Yes}           \\

		\makecell{Primary        \\Goal} &
		\makecell{Visualize      \\Semantics} &
		\makecell{Fault          \\Localization} &
		\makecell{Visualize      \\State/Refs} &
		\makecell{Beginner       \\Debugging} &
		\makecell{Production     \\Debugging} \\
		\bottomrule
	\end{tabular}
\end{table}

\section{Limitations and Future Work}

While the SimpliPy notional machine and its associated visualization tool
provide a focused pedagogical foundation for understanding Python's control flow
and scope, certain limitations exist in the current implementation, suggesting
avenues for future development.

One key limitation is the restricted language subset, omitting features like
exceptions, comprehensions, modules, asynchronous programming, and objects.
Future work involves carefully extending the semantics for such features,
balancing coverage with pedagogical clarity.

The visualization tool, while effective for observation, currently lacks
interactive debugging capabilities common in production environments, such as
setting breakpoints or modifying variable values during execution. Future work
could explore adding limited, pedagogically motivated interactive features,
alongside enhanced visualizations focusing on data flow.

Finally, rigorous empirical studies are required to formally evaluate the
effectiveness of the SimpliPy approach. Future work includes conducting
controlled user studies with introductory programming students to compare
learning outcomes (particularly regarding mental model accuracy and
misconception reduction) when using SimpliPy (both manual tracing and the tool)
versus traditional teaching methods or other visualization tools.

\section{Conclusion}

SimpliPy presents a formally grounded approach to teaching fundamental aspects
of Python program execution, specifically control flow and lexical scoping. By
defining a precise operational semantics for a carefully curated language subset
with explicit source-tracking, integrating static analysis artifacts like
Control Flow Graphs and lexical scope information, and providing an interactive
visualization tool, it aims to help novice programmers build more accurate and
robust mental models. The emphasis on comprehension prior to tracing,
facilitated by static analysis, and the subsequent reinforcement via the
visualization tool which faithfully renders the notional machine's state
(environments, lexical hierarchy, continuation) and control flow, offers a
distinct pedagogical pathway.

The SimpliPy tool, in particular, demonstrates the value of applying formal
methods in an educational context, providing a level of abstraction over
low-level memory details (stack frames, heap) that allows learners to focus
directly on semantic rules. While intentionally simpler than production
debuggers or broader educational visualizers, its strength lies in this focused,
semantic-driven visualization. SimpliPy provides a solid foundation for both
teaching core programming concepts and further research into semantics-based
programming education.

The supplementary materials repository, available at
\url{https://github.com/PraneethJain/simplipy}, contains the detailed resources
referenced in this paper, including the formal BNF grammar, the complete
operational semantics transition rules, and the implementation of the SimpliPy
visualization tool.

\bibliographystyle{splncs04}
\bibliography{bib/misc/ref.bib
	,bib/misconceptions/ref.bib
	,bib/notional-machines/ref.bib
	,bib/prog-languages-and-python/ref.bib
	,bib/tracing-program-comprehension/ref.bib
}

\end{document}